**The influence of the distribution function of ferroelectric nanoparticles sizes on their electrocaloric and pyroelectric properties**

***Hanna V. Shevliakova*[1], *Anna N. Morozovska* [2*], *Nicholas V. Morozosky*[2], *George S. Svechnikov*[1] and *Vladimir V. Shvartsman*[3]**

[1] *Department of Microelectronics, National Technical University of Ukraine "Igor Sikorsky Kyiv Polytechnic Institute", Kyiv, Ukraine*

[2] *Institute of Physics, National Academy of Sciences of Ukraine, 46, pr. Nauky, 03028 Kyiv, Ukraine*

[3]*Institute for Materials Science and Center for Nanointegration Duisburg-Essen (CENIDE), University of Duisburg-Essen, 45141, Essen, Germany*

We consider a model of a nanocomposite based on non-interacting spherical single-domain ferroelectric nanoparticles of various sizes embedded in a dielectric matrix. The size distribution function of these nanoparticles is selected as a part of the Gaussian distribution from minimum to maximum radius (truncated normal distribution). For such nanocomposites, we calculate the dependences of the reversible part of the electric polarization, the electrocaloric temperature change, and the dielectric permittivity on the external electric field, which have the characteristic form of hysteresis loops. We then analyze the change in the shape of the hysteresis loops relative to the particle size distribution parameters. We demonstrate that for the same mean-square dispersion, the remanent polarization, coercive field, dielectric permittivity maximums, maximums and minimums of the electrocaloric temperature change depend most strongly on the most probable radius, moderately depend on the dispersion, and have the weakest dependency on the nanoparticle maximum radius. We calculated and analyzed the dependences of pyroelectric figures of merit on the average radius of the nanoparticles in the composite. The dependences confirm the presence of a phase transition induced by the size of the nanoparticles, which is characterized by the presence of a maxima near the critical average radius of the particles, the value of which increases with increasing dispersion of the distribution function.

[*] corresponding author, e-mail: anna.n.morozovska@gmail.com

## I.Introduction

From the second half of the 20th century to the present, ferroelectric (FE) materials have been the objects of intense experimental and theoretical studies due to their use as active media in a number of converting devices, in particular in pyroelectric (PE) [1, 2, 3] and electrocaloric (EC) [1, 2, 4, 5] converters. For many years, pyroelectric converters have been used in many applications from gas detectors to thermal imaging [6], however, only the recently discovered “giant” EC effect in thin films [7] opened up the prospect for using the EC effect in solid-state microcoolers. The PE and EC properties of thin ferroelectric films, multilayers, and other low-dimensional materials can differ greatly from those of bulk materials. In particular, the prospects of using FE nanocomposites for EC converters [8, 9, 10] and PE sensors [11] are more compelling. Therefore, studies of low-dimensional FE materials, such as thin films and nanocomposites, are very relevant [3, 5, 11, 12, 13]. The study of EC cooling is of great importance to finding solutions to environmental problems [5, 12] and energy efficiency [14] of currently available cooling technologies.

Further progress in this direction is hindered by a number of technological and theoretical difficulties [15, 16]. These difficulties relate to the appearance of a practically unremovable electric field of depolarization, which is not taken into account when considering EC and PE effects [17].

Modern methods allow precise selection of nanoparticles by size and shape, however, nanocomposites made on their basis, as a rule, contain nanoparticles with a more or less symmetric distribution in size within certain limits around the average size [18, 19, 20].

As indicated in [21], it is still unclear what effect the size distribution of ferroelectric nanoparticles has on the EC properties of nanocomposites based on them. In his case, the properties of the composite depend on the predominance of the contribution of particles of one size or another. The numerical and analytical models developed to date are mainly aimed at the description of composites with nanoparticles of the same size and certain shape [8, 22, 23, 24].

This article is essentially a semi-analytical and semi-numerical description of the EC and PE properties of nanocomposites based on ferroelectric nanoparticles with the most realistic Gaussian size distribution function.

## II.Problem Statement

We consider a nanocomposite consisting of an isotropic dielectric matrix with permittivity $\varepsilon_e$ and immersed ferroelectric nanoparticles with permittivity $\varepsilon_b$. Each ferroelectric nanoparticle is surrounded by a semiconductor shell with a dielectric constant $\varepsilon_{IF}$, which acts as a layer screening the ferroelectric polarization of a particle with a thickness equal to the “effective” screening length $\Lambda$ [25]. The spread of the radii of the nanoparticle sizes is in a range from minimum $R_{min}$ to maximum $R_{max}$.

A schematic representation of the model of the nanocomposite under consideration is shown in **Fig. 1**. Due to the screening, the interaction between the particles in a nanocomposite can be neglected if the relative fraction of the volume of the nanoparticles is small (less than 10%). However, we note that if the degree of screening is very high, the interaction between the nanoparticles disappears, and the interaction of the nanoparticles with an external electric field is weakened. It is believed that the degree of screening is independent on the particle concentration, which is true up to very high concentrations.

Ferroelectric nanoparticles were previously polarized by a strong electric field while the polymer was in the liquid phase and the particles could rotate almost freely in it. At that the Curie temperature of ferroelectric nanoparticles should be significantly higher than the polymer melting temperature, and the poling field should be significantly smaller than the breakdown field of the liquid polymer. After polymer solidification, it can be assumed that all nanoparticles are single domain with the only component of spontaneous polarization $P_3(\mathbf{r})$ directed along axis 3 of the perovskite unit cell.

The model structure of the core-shell nanoparticle $BaTiO_3$ under consideration is in accordance with the X-rays synchrotron radiation analysis [26] and scanning transmission electron microscopy observation [27] data, indicating the presence of an inner tetragonal core, gradient lattice strain layer, and surface cubic layer [28], which was used earlier [8, 28] to evaluate the efficiency of EC conversion of these nanoparticles.

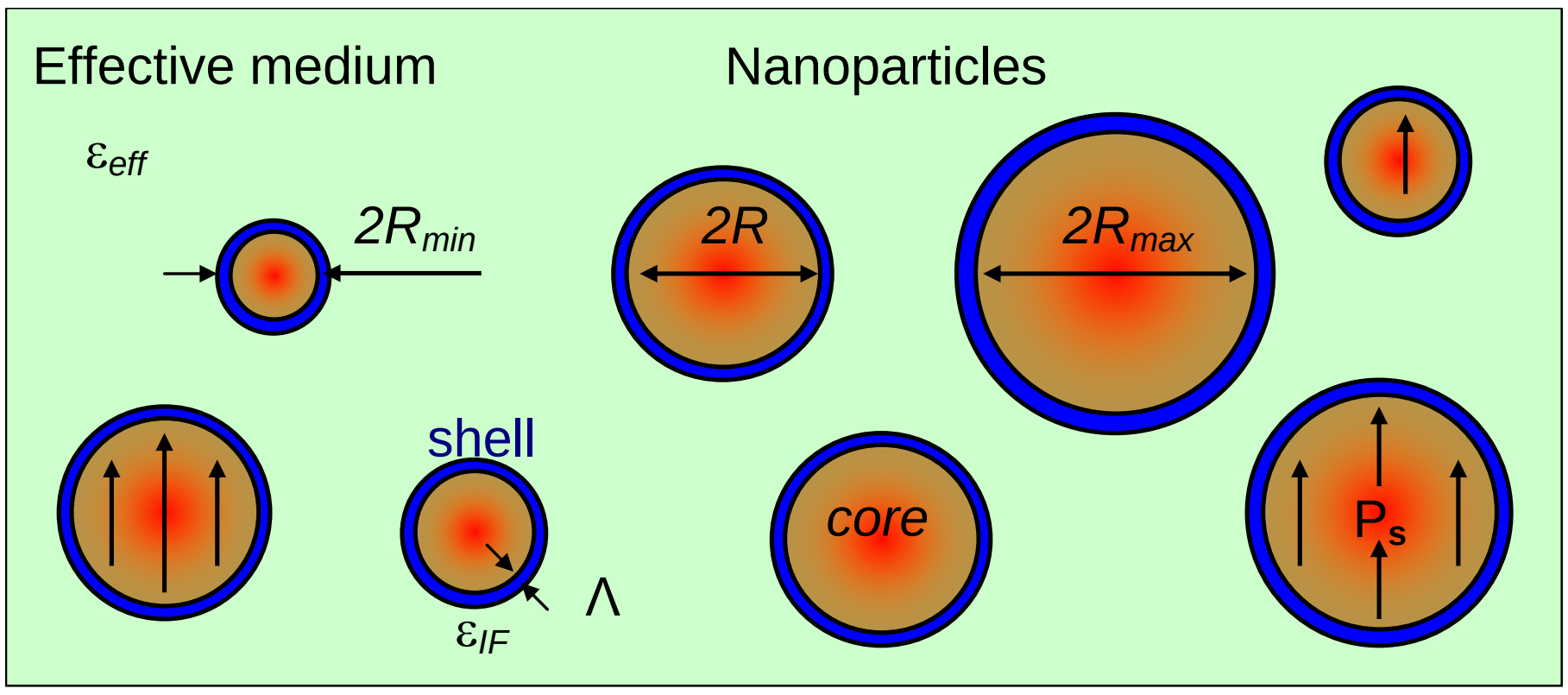

**FIGURE 1.** Spherical ferroelectric nanoparticles of different radii covered with a thin semiconducting shell and placed in an isotropic dielectric polymer.

For calculations, we assume that the radii distribution of the nanoparticles corresponds to a distribution function $f(R)$, which is expressed by the normal Gaussian distribution:

$$f(R)=\frac{1}{\sigma_R}\exp\left[-\frac{(R-R_m)^2}{2\sigma^2}\right], \tag{1}$$

where $\sigma^2$ is the dispersion characterizing the spread of $R$ around the most probable radius $R_m$, and $\sigma_R$ is the normalizing coefficient. Given that the particle radii vary from $R_{min}$ to $R_{max}$, the normalization condition is ~~satisfied~~:

$$\int_{R_{\min}}^{R_{\max}} f(R)\,dR = 1, \tag{2a}$$

where the normalizing coefficient $\sigma_R$ is

$$\sigma_R = \sqrt{\frac{\pi}{2}}\sigma\left[\operatorname{erf}\left(\frac{R_m - R_{\min}}{\sqrt{2}\sigma}\right) - \operatorname{erf}\left(\frac{R_m - R_{\max}}{\sqrt{2}\sigma}\right)\right], \tag{2b}$$

radii $R_{\min} \le R_m \le R_{\max}$, and $\operatorname{erf}(x) = \frac{2}{\sqrt{\pi}}\int_0^x \exp(-z^2)\,dz$ is error function.

In mathematics, the parameter $\sigma > 0$ represents the normal deviation, however, in the physics literature, both quantities, $\sigma^2$ and $\sigma$, often represent dispersion, despite the different dimensions. Below, we will denote $\sigma > 0$ the dispersion for simplicity.

The average radius is calculated by the formula

$$\bar{R}(R_m) = \int_{R_{\min}}^{R_{\max}} R \cdot f(R)\,dR = \frac{\sigma^2}{\sigma_R}\left(\exp\left[-\left(\frac{R_m - R_{\min}}{\sqrt{2}\sigma}\right)^2\right] - \exp\left[-\left(\frac{R_m - R_{\max}}{\sqrt{2}\sigma}\right)^2\right]\right) + R_m . \tag{2c}$$

and differs from $R_m$ as the gaussoid is "cut off" in the range from $R_{min}$ to $R_{max}$.

The dependence of the distribution function $f(R)$ on the parameters $R_m$, $\sigma$ and $R_{max}$ is presented in **Figs. 2a, 3a** and **4a**, respectively, where the same $R$ interval was used for ease of comparison.

Earlier we analyzed the typical dependences of the PE parameters and EC conversion on the external electric field $E_{ext}$, temperature, and radius of spherical single-domain FE nanoparticles with a fixed radius [22], using the phenomenological Landau-Ginsburg-Devonshire (LGD) theory and effective medium approximation.

It should be noted that the "bulk" LGD-coefficients, renormalized by size effects, can be used to describe the spatially confined ferroelectric micro- and nanosystems [22, 25]. Contributions of strains and polarization gradients, as well as the depolarization and screening effects were taken into consideration by introducing the appropriate factor depending on values of the relative dielectric permittivity of the nanoparticles ($\varepsilon_b$), shell ($\varepsilon_{IF}$) and surrounding medium ($\varepsilon_e$), nanoparticle radius $R$ and "effective" screening length $\Lambda$ [22]:

$$\eta(R,\Lambda) = \frac{3\varepsilon_e}{\varepsilon_b + 2\varepsilon_e + \varepsilon_{IF}(R/\Lambda)} . \tag{3}$$

The approximate expression for the nanoparticle transition temperature $T_{cr}$ from the single-domain ferroelectric to the paraelectric phase is [22]

$$T_{cr}(R,\Lambda)=T_C^*-\frac{\eta(R,\Lambda)}{3\alpha_T\varepsilon_0\varepsilon_e}. \tag{4}$$

where the first term $T_C^*$ is Curie temperature (possibly renormalized by the surface stress [29]) and $\alpha_T$ is the inverse Curie-Weiss constant. The second term originates from a depolarization field. Polarization obeys the time-dependent LGD equation [25, 22]

$$\Gamma\frac{\partial P}{\partial t}+\alpha_T(T-T_{cr})P+\beta P^3+\gamma P^5-g_{44}\left(\frac{\partial^2}{\partial x_1^2}+\frac{\partial^2}{\partial x_2^2}\right)P-g_{11}\frac{\partial^2 P}{\partial x_3^2}=\eta E_{ext}, \tag{5}$$

where $\Gamma$ is the Khalatnikov's kinetic coefficient, and $g_{11}$ and $g_{44}$ are the gradient coefficients.

For $BaTiO_3$ the coefficients $\alpha$, $\beta$ and $\gamma$ are temperature-dependent in a similar linear way as $\alpha=\alpha_T(T-T_{cr})$, $\beta=\beta_T(T-T_\beta)$ and $\gamma=\gamma_T(T-T_\gamma)$. Coefficient $\beta$ is negative in the considered case of the 1$^{st}$ order ferroelectric phase transition. Subsequently, positive gradient coefficients $g_{44}$ and $g_{11}$ are regarded either small enough, or already included in the renormalization of $T_C^*$. This allows us to ignore the last two gradient terms in Eq.(5).

The dynamic dielectric susceptibility, defined as $\chi_{33}=\frac{\partial P}{\partial E_{ext}}$, obeys the equation [22]:

$$\Gamma\frac{\partial\chi_{33}}{\partial t}+\left[\alpha_T(T-T_{cr}(R,\Lambda))+3\beta P^2+5\gamma P^4\right]\chi_{33}=\eta \tag{6}$$

Differentiation of the static equation (5) with respect to temperature leads to the equation $\left(\frac{\partial P}{\partial T}\right)_E\left[\alpha_T(T-T_{cr})+3\beta P^2+5\gamma P^4\right]=-\alpha_T P-\beta_T P^3-\gamma_T P^5$. Using this equation, the analytical expression for the PE coefficient is:

$$\Pi(R,\Lambda)=\frac{\alpha_T P+\beta_T P^3+\gamma_T P^5}{\alpha_T\left[T-T_{cr}(R,\Lambda)\right]+3\beta P^2+5\gamma P^4}, \tag{7}$$

In the case of a ferroelectric with the linear temperature dependence of coefficient $\alpha$ in LGD-expansion (5) (e.g. for $BaTiO_3$), the EC temperature change $\Delta T_{EC}$ can be calculated from the expression [22]:

$$\Delta T_{EC}=\frac{T}{\rho}\int_0^E\frac{\Pi(R,\Lambda,P(E))}{C_P}dE\approx\frac{T}{\eta\rho C_P}\left(\frac{\alpha_T}{2}\left[P^2(E)-P^2(0)\right]+\right.$$
$$\left.+\frac{\beta_T}{4}\left[P^4(E)-P^4(0)\right]+\frac{\gamma_T}{6}\left[P^6(E)-P^6(0)\right]\right) \tag{8a}$$

Since the nanocomposite contains nanoparticles of different sizes, the required parameters should be averaged with the distribution function $f(R)$:

$$\Delta T_{EC}\left(\Lambda, E_{ext}\right) = \int_{R_{\min}}^{R_{\max}} \Delta T_{EC}\left(R, \Lambda, E_{ext}\right) f\left(R\right) dR, \tag{8b}$$

where $\Delta T_{EC}\left(R, \Lambda, E_{ext}\right)$ is given by Eq.(8a).

The EC coefficient $\Sigma\left(E_{ext}\right)$ is defined as the derivative of the EC temperature change $\Delta T_{EC}(E_{ext})$ with respect to the external electric field:

$$\Sigma = \frac{d\Delta T_{EC}}{dE_{ext}}. \tag{9}$$

Relative dielectric permittivity, $\varepsilon_{33} = 1 + \frac{\chi_{33}}{\varepsilon_0}$, for the static or very low frequency dynamic case is

$$\varepsilon_{33} \approx 1 + \frac{\eta(R,\Lambda)}{\varepsilon_0\left[\alpha_T\left(T - T_{cr}(R,\Lambda)\right) + 3\beta P^2 + 5\gamma P^4 + i\omega\Gamma\right]}, \tag{10a}$$

$$\varepsilon_{33}\left(\Lambda, E_{ext}\right) = \int_{R_{\min}}^{R_{\max}} \varepsilon_{33}\left(R, \Lambda, E_{ext}\right) f\left(R\right) dR. \tag{10b}$$

The heat capacity is [22]:

$$C_P = C_P^0 + \delta C_P, \tag{11a}$$

$$\delta C_P = \frac{T\left(\alpha_T P + \beta_T P^3 + \gamma_T P^5\right)^2}{\alpha_T\left(T - T_{cr}\right) + 3\beta P^2 + 5\gamma P^4} = T\left(\alpha_T P + \beta_T P^3 + \gamma_T P^5\right)\Pi\left(R,\Lambda\right). \tag{11b}$$

LGD parameters for bulk ferroelectric $\mathrm{BaTiO_3}$ are given in **Table 1**. The critical radius of the size induced ferroelectric-paraelectric phase transition, $R_{cr} \approx 8$ nm, was calculated in Ref.[30].

**Table 1.** LGD parameters for bulk ferroelectric $\mathrm{BaTiO_3}$*

| Parameters | Value |
|---|---|
| $\varepsilon_b$ | 7 |
| $\alpha_T\left(\mathrm{C^{-2}\cdot m\cdot J/K}\right)$ | $6.68\cdot 10^5$ |
| $T_C\left(\mathrm{K}\right)$ | 381 |
| $\beta\left(\mathrm{C^{-4}\cdot m^5\cdot J}\right)$ | $\beta_T\cdot(T-393) - 8.08\cdot 10^8$, $\beta_T = 18.76\cdot 10^6$ |
| $\gamma\left(\mathrm{C^{-6}\cdot m^9\cdot J}\right)$ | $\gamma_T\cdot(T-393) + 16.56\cdot 10^9$, $\gamma_T = -33.12\cdot 10^7$ ** |

* $\rho = 6.02\cdot 10^3\ \mathrm{kg/m^3}$, $c_p^0 = 4.6\cdot 10^2\ \mathrm{J/(kg\cdot K)}$ and so $C_p^0 = \rho c_p^0$ in $J/\left(m^3\cdot K\right)$ at room temperature.

** These parameters are valid until $\gamma > 0$, i.e. for $T < 445$ K.

## III. Results and discussion

### A. Polarization hysteresis loops, EC temperature changes and dielectric permittivity

The obtained dependences of polarization $P$, EC on temperature change $\Delta T_{EC}$ and dielectric permittivity $\varepsilon_{NP}$ on an external electric field $E_{ext}$ shown in **Figs. 2b-d, 3b-d** and **4b-d** have the form of hysteresis loops.

Hysteresis loops $P(E)$, $\Delta T_{EC}(E)$, and $\varepsilon_{NP}(E)$, shown in **Figs. 2b-d**, correspond to different values of the most probable $R_m$ of distribution function $f(R)$, varying in the interval 5 $nm \leq R_m \leq$ 17 $nm$ (see **Fig. 2a**). Other parameters of $f(R)$ were fixed at $R_{min} = 1$ nm, $R_{max} = 40$ nm and $\sigma = 5$ nm.

With such a change in parameters of $f(R)$, with decreasing $R_m$, the number of particles with a radius $R < R_m$ decreases, and the number of particles with $R > R_m$ practically does not change (compare the curves 1-4 in **Fig. 2a**). Note that different $\bar{R}$ correspond to different $R_m$, namely $\bar{R}$ = 6.84, 9.59, 13.11 and 17.01 nm for $R_m$ = 5, 9, 13 and 17 nm. Therefore, in **Fig. 2a**, the inscriptions for the same curves 1–4 indicate both quantities, $\bar{R}$ and $R_m$.

We immediately note that the characteristic features of the hysteresis loops associated with the proximity of the most probable particle radius to the critical radius $R_{cr} \approx 8$ nm, are best observed for $R_m = 9$ nm on loops $\varepsilon_{NP}(E)$ (see red loops 2). When $R_m = 5$ nm most particles are in the paraelectric phase, and when $R_m$ =17 nm – in the FE phase.

An increase in $R_m$, leads to a decrease in the average slope of the polarization hysteresis loop $P(E)$, an increase in the remanent polarization $P_r$ and coercive field $E_c$, and also to a slight decrease in the maximum polarization $P_{max}$ (see **Table 2** and **Fig. 2b**).

In this case, the $\Delta T_{EC}(E)$ loop is deformed in such a way that the negative maxima $\Delta T_{EC}$ expand near $E_c$ and their absolute value $\Delta T_{EC}^{max}$ increases, while the positive value $\Delta T_{EC}$ on the "shoulders" of the $\Delta T_{EC}(E)$ loop decreases (see **Fig. 2c** and **Table 2**).

With an increase in $R_m$, the height of the $\varepsilon_{NP}(E)$ loop maxima near $E_c$ monotonously increases (see **Fig. 2d** and **Table 2**), and the expansion of the maxima is similar to the expansion of the maxima $\Delta T_{EC}$, and corresponds to a decrease in the slope of the $P(E)$ loop in **Fig. 2b.**

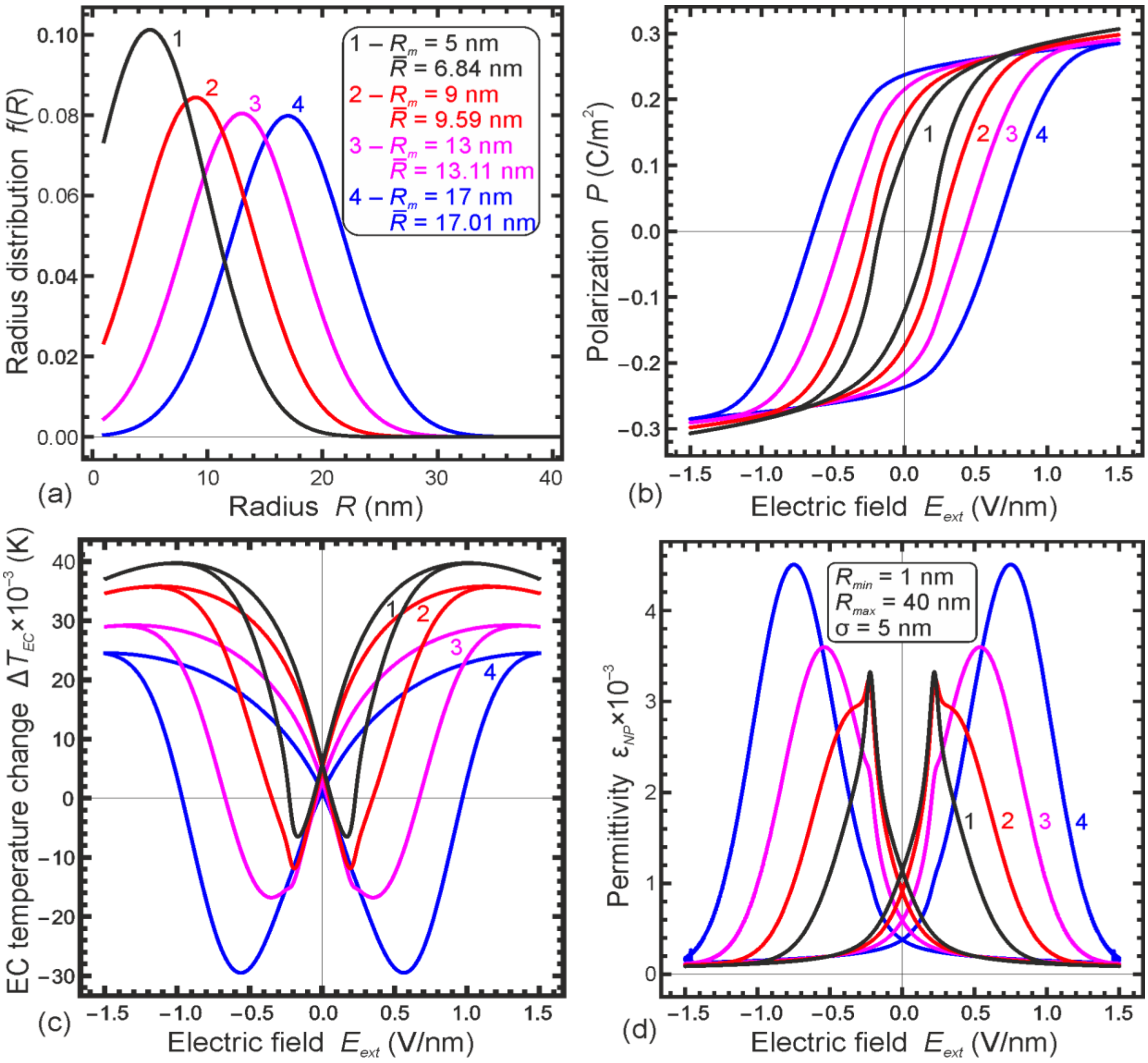

**FIGURE 2. (a)** Distribution functions of nanoparticle radii calculated for different parameter $R_m$ = 5, 9, 13, 17 nm (curves 1-4), fixed $R_{min} = 1$ nm, $R_{max} = 40$ nm and $\sigma = 5$ nm. Dependences of the polarization **(b)**, EC temperature change **(c)** and relative dielectric permittivity **(d)** on external electric field calculated for an ensemble of noninteracting $BaTiO_3$ nanoparticles (curves 1-4), distributed in accordance with figure (a), $T = 293\,K$, $\varepsilon_{IF} = 300$, $L = 2$ nm, $\varepsilon_e = 15$, $\Gamma = 10^2$ Ohm/m, and $\omega = 2\cdot 10^4\,s^{-1}$. $BaTiO_3$ parameters are listed in **Table 1**.

**Table 2.** Parameters of hysteresis loops $P(E)$, $\Delta T_{EC}$ and $\varepsilon_{NP}(E)$ at fixed $R_{min} = 1$ nm, $R_{max} = 40$ nm and $\sigma = 5$ nm.

| $R_m$, nm | 5 | 9 | 13 | 17 |
|---|---|---|---|---|
| $P_{\max}$, $\mathrm{C/m^2}$ | 0.305 | 0.3 | 0.29 | 0.285 |
| $P_r$, $\mathrm{C/m^2}$ | 0.12 | 0.18 | 0.22 | 0.24 |
| $E_c$, V/nm | 0.15 | 0.25 | 0.4 | 0.65 |
| $\Delta T_{EC}{}^{max}\cdot 10^{-3}$, K | -6 | -12 | -17 | -29 |
| $\Delta T_{EC}{}^{should}\cdot 10^{-3}$, K | 40 | 36 | 30 | 25 |
| $\varepsilon_{NP}{}^{max}\cdot 10^{-3}$ | ≈2.7 | ≈3 | 3.6 | 4.5 |

The hysteresis loops $P(E)$, $\Delta T_{EC}(E)$ and $\varepsilon_{NP}(E)$, shown in **Figs. 3b-d**, correspond to different dispersion values $\sigma$ of the distribution function $f(R)$, varying in the interval $1\,nm \leq \sigma \leq 7\,nm$ (see **Fig. 3a**). Other parameters $f(R)$ were fixed at: $R_{min} = 1\,nm$, $R_m = 5\,nm$ and $R_{max} = 40\,nm$. Note that different $\sigma$ correspond to different $\bar{R}$ values, namely $\bar{R} = 8.31$, 6.84, 5.54 and 5.00 nm for $\sigma = 7$, 5, 3, and 1 nm. Therefore, in **Fig. 3a**, the inscriptions for the same curves 1–4 indicate both values, $\bar{R}$ and $\sigma$.

With a decreasing $\sigma$, function $f(R)$ becomes much better localized near the maximum at $R = R_m$. Since $R_m$ is smaller than the critical radius, there are features for blue loops 4 associated with most of the particles in the composite being in the paraelectric phase. This is why a decrease in $\sigma$ leads to a decrease in the remanent polarization $P_r$ and the coercive field $E_c$ with a slight change in the maximum polarization $P_{max}$ and an increase in the average slope of the narrow $P(E)$ loop, characteristic of small $R_m < R_{cr}$ (see **Table 3** and **Fig. 3b**). In this case, the $\Delta T_{EC}(E)$ loop is deformed in such a way that the negative maxima of $\Delta T_{EC}$ near $E_c$ become narrow with a decrease in their absolute value $\Delta T_{EC}{}^{max}$ and change sign, and the positive value $\Delta T_{EC}$ on the “shoulders” of the $\Delta T_{EC}(E)$ loop increases (see **Table 3** and **Fig. 3c**).

With decreasing $\sigma$, the height of the $\varepsilon_{NP}(E)$ loop maxima near $E_c$ changes non-monotonously (see **Table 3** and **Fig. 3d**), and the narrowing of the $\varepsilon_{NP}(E)$ maxima is similar to the narrowing of the $\Delta T_{EC}(E)$ maxima, and corresponds to an increase in the slope of the $P(E)$ loop in **Fig. 3b**. The central maximum on the $\varepsilon_{NP}(E)$ loop 4 appears as most of the particles in the composite are in the paraelectric phase, and the lateral maximums correspond to the fraction of particles in the ferroelectric phase.

**Table 3.** Parameters of hysteresis loops $P(E)$, $\Delta T_{EC}(E)$ and $\varepsilon_{NP}(E)$ at fixed $R_{min} = 1$ nm, $R_{max} = 40$ nm and $R_m = 5$ nm.

| σ, nm | 1 | 3 | 5 | 7 |
|---|---|---|---|---|
| $P_r$, $\mathrm{C/m^2}$ | 0.05 | 0.09 | 0.12 | 0.15 |
| $E_c$, V/nm | 0.08 | 0.12 | 0.17 | 0.2 |
| $\Delta T_{EC}^{max} \cdot 10^{-3}$ K | +1 | -3.5 | -6.5 | -8 |
| $\Delta T_{EC}^{should} \cdot 10^{-3}$ K | 52 | 44 | 40 | 36 |
| $\varepsilon_{NP}^{max} \cdot 10^{-3}$ | ≈3 | 3.5 | 3.3 | ≈3 |

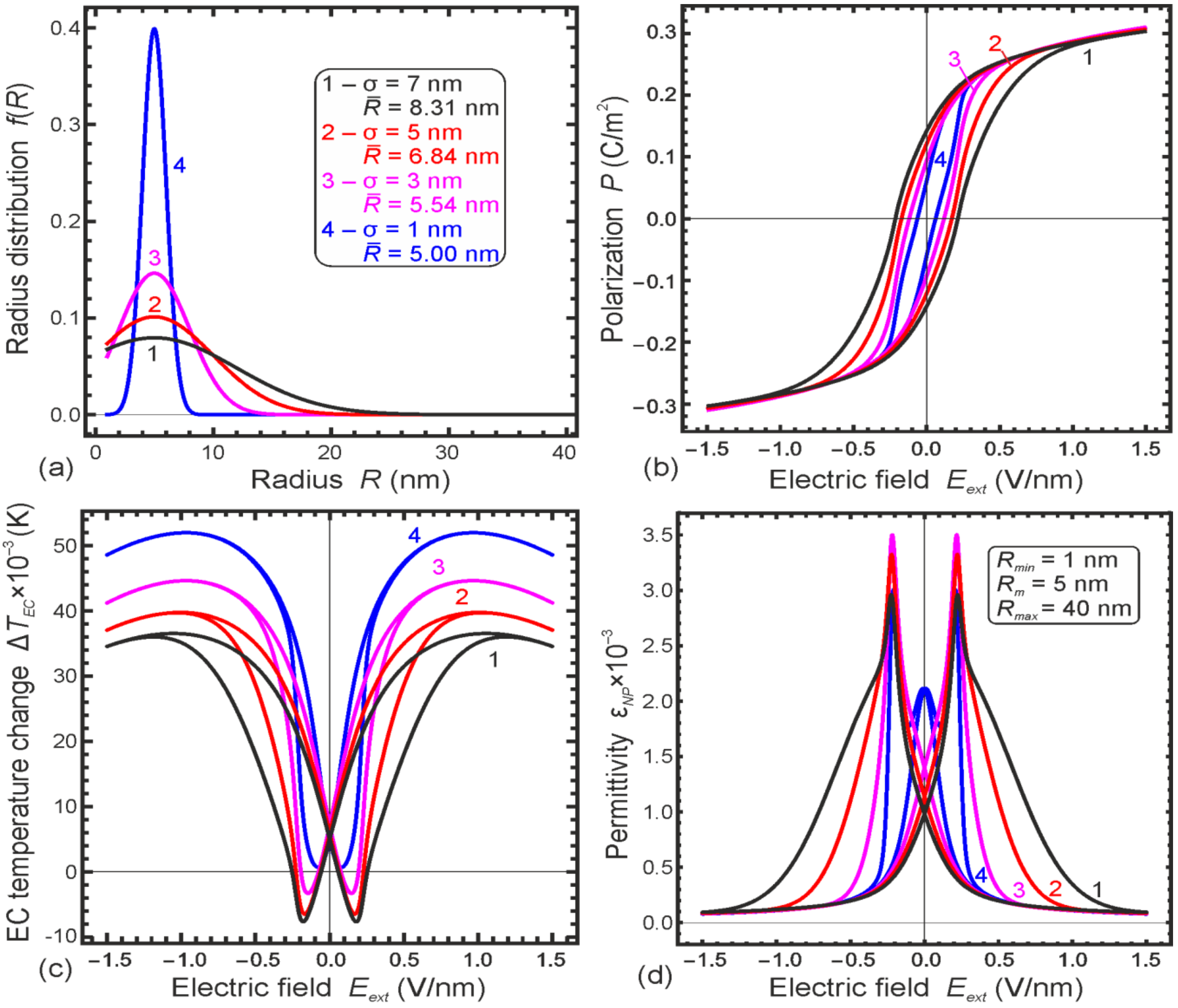

**FIGURE 3. (a)** Distribution functions of nanoparticle radii calculated for different dispersion $\sigma =$ 7, 5, 3, and 1 nm (curves 1-4), fixed $R_{min} = 1$ nm, $R_{max} = 40$ nm and $R_m = 5$ nm. Dependences of the polarization **(b)**, EC temperature change **(c)** and relative dielectric permittivity **(d)** on external electric field calculated for an ensemble of noninteracting $BaTiO_3$ nanoparticles (curves 1-4), distributed in accordance with figure **(a)**. Other parameters are the same as in **Fig. 2**.

Hysteresis loops $P(E)$, $\Delta T_{EC}(E)$ and $\varepsilon_{NP}(E)$, shown in **Figs. 4b-d**, correspond to different values of the maximal radius $R_{max}$ of the distribution function $f(R)$, varying in the interval $10\ nm \leq R_{max} \leq 25\ nm$ (see **Fig. 4a**). Other parameters of $f(R)$ were fixed at: $R_{min} = 1\ nm$, $R_m = 5\ nm$ and $\sigma = 5\ nm$. At that, with a $R_{max}$ decrease, the ratio of the number of particles with different deviations $R$ from $R_m$ changes in favor of particles with $R_{max}$ close to $R_m$. Note that different $R_{max}$ correspond to different values $\bar{R}$, namely $\bar{R} = 6.84$, 6.81, 6.54 and 5.38 nm for $R_{max} = 25$, 20, 15 and 10 nm. Therefore, in **Fig. 4a**, in the inscriptions for the same curves 1–4, both quantities, $\bar{R}$ and $R_{max}$, are indicated.

A decrease in $R_{max}$ leads to an increase in the slope of the $P(E)$ hysteresis loop, a slight decrease in the remanent polarization $P_r$ and coercive field $E_c$ ,while maintaining the maximum polarization $P_{max}$ (see **Table 3** and **Fig. 4b**). The shape of $\Delta T_{EC}(E)$ loop, the negative maximum of $\Delta T_{EC}$ near $E_c$ , and their absolute value $\Delta T_{EC}^{max}$ , as well as the positive value $\Delta T_{EC}$ on the "shoulders" of $\Delta T_{EC}(E)$, vary slightly (see **Table 3** and **Fig. 4c**). With a decrease in $R_{max}$, the height of the $\varepsilon_{NP}(E)$ loop maxima near EC increases slightly (see **Table 3** and **Fig. 4d**), and the narrowing of the $\varepsilon_{NP}$ maxima is similar to the narrowing of the $\Delta T_{EC}$ maxima, and corresponds to a change in the shape of the $P(E)$ loop in **Fig. 4b**. Generally speaking, the evident conclusion follows from **Fig. 4**: the particles with a radius $R_{max} > 3\sigma$ practically does not contribute to the properties of the nanocomposite.

**Table 4.** Parameters of hysteresis loops $P(E)$, $\Delta T_{EC}(E)$ and $\varepsilon_{NP}(E)$ at fixed $R_{min} = 1$ nm, $R_m = 5$ nm and $\sigma = 5$ nm.

| $R_{\max}$, nm | **10** | **15** | **20** | **25** |
|---|---|---|---|---|
| $P_r$, C/m$^2$ | 0.1 | 0.12 | 0.12 | 0.12 |
| $E_c$, V/nm | 0.12 | 0.18 | 0.18 | 0.18 |
| $\Delta T_{EC}^{max} \cdot 10^{-3}$, K | -4 | -6 | -6 | -6 |
| $\Delta T_{EC}^{should} \cdot 10^{-3}$, K | 44 | 40 | 40 | 40 |
| $\varepsilon_{NP}^{max} \cdot 10^{-3}$ | ≈3.8 | ≈3.4 | 3.4 | 3.4 |

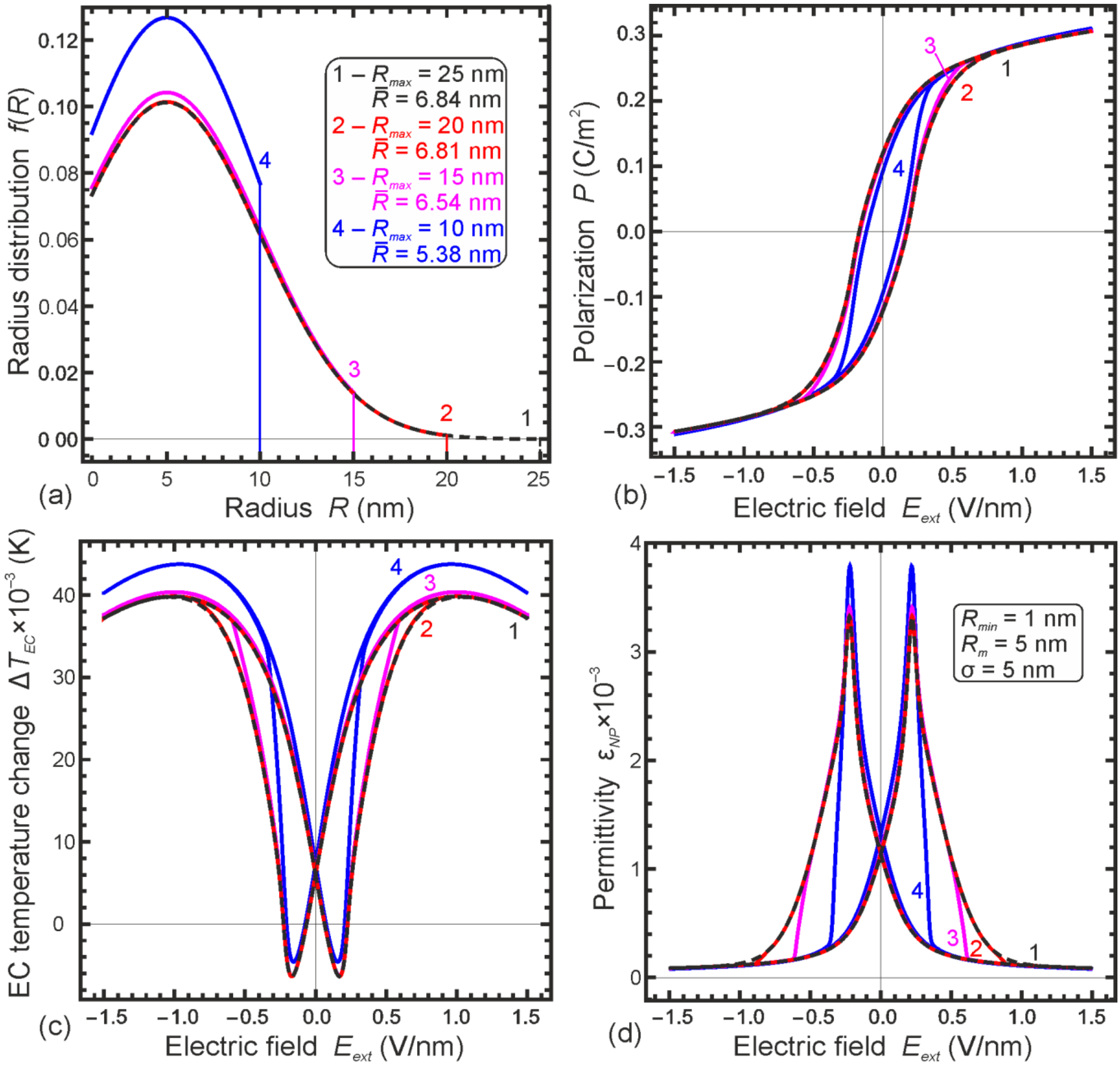

**FIGURE 4. (a)** Distribution functions of nanoparticle radii calculated for different maximal radius $R_{max} = 25$, 20, 15, and 10 nm (curves 1-4), fixed $R_{min} = 1$ nm, $R_m = 5$ nm and $\sigma = 5\ nm$. Dependences of the polarization **(b)**, EC temperature change **(c)** and relative dielectric permittivity **(d)** on external electric field calculated for an ensemble of noninteracting $BaTiO_3$ nanoparticles (curves 1-4), distributed in accordance with figure **(a).** Other parameters are the same as in **Fig. 2**.

In summary, we calculated and analyzed the changes in the shape of the hysteresis loops $P(E)$, $\Delta T_{EC}(E)$ and $\varepsilon_{NP}(E)$ originated from the change of parameters of the nanoparticle size distribution function: the most probable and maximum radii, and dispersion (standard deviation) with the constant minimum radius of 1 nm. We have demonstrated that for the same standard deviation (5 nm), the remanent polarization, the coercive field, the dielectric permittivity maximums and the negative maximums of the EC temperature change depend substantially on the most probable radius [in the range of (5 – 17) nm] and weakly depend on the maximum radius [in the range of (10 – 25) nm]. For particles with the small most probable (5 nm) and large maximum (40 nm) radii, the above values decrease with decreasing standard deviation in the range of (1 – 5 )nm.

### B. Correlation of the shape and characteristic features of EC and PE hysteresis

The dependence of the pyroelectric $\Pi$ and electrocaloric $\Sigma$ coefficients on the external electric field $E_{ext}$ , are shown in **Figs. 5** and have the form of hysteresis loops. According to the field dependencies of $P(E)$ and $\Delta T_{EC}(E)$ presented in **Figs. 2-4,** $\Pi$ and $\Sigma$ hysteresis loops are symmetrical with respect to the zero point. The shape of hysteresis loops and $E_c$ values depend on the parameters of the particle radius distribution function ($R_m$, $\sigma$ and $R_{max}$) in accordance with Eqs. (7)-(9).

The loops in **Figs. 5a, b** correspond to different $R_m$ in the distribution functions $f(R)$, varying in the interval $5\ nm \le R_m \le 17\ nm$, and fixed $R_{min} = 1$ nm, $R_{max} = 40$ nm and $\sigma = 5$ nm (see **Fig. 2a**). With this change in the distribution of particles radii, both the PE and EC coefficients are characterized by the presence of double maxima, which increase with increasing $R_m$, and expand and shift towards the large fields (see curves 2-4). When the radius $R_m = 5\ nm$, that is less than the critical value $R_{cr} \approx 8$ nm , the appearance of an additional maximum in both dependences $\Pi(E_{ext})$ and $\Sigma(E_{ext})$ is observed (see curves 1). The double maxima at both the dependences $\Pi(E_{ext})$ and $\Sigma(E_{ext})$ for Rm = 5 nm (curves 1) are related to the peculiarities of the dependences $P(E)$ and $\Delta T_{EC}(E)$ at the balance of particles with $R < R_{cr}$ and $R > R_{cr}$ at $\sigma$ = 5nm (see curves 1 in Fig. 2a, b, c). As $R_m$ increases, the dependences $\Pi(E_{ext})$ and $\Sigma(E_{ext})$ are characterized by the existence of maxima, which, moving towards the larger $E_{ext}$ (that also correspond to the increase of $E_c$), increase and expand.

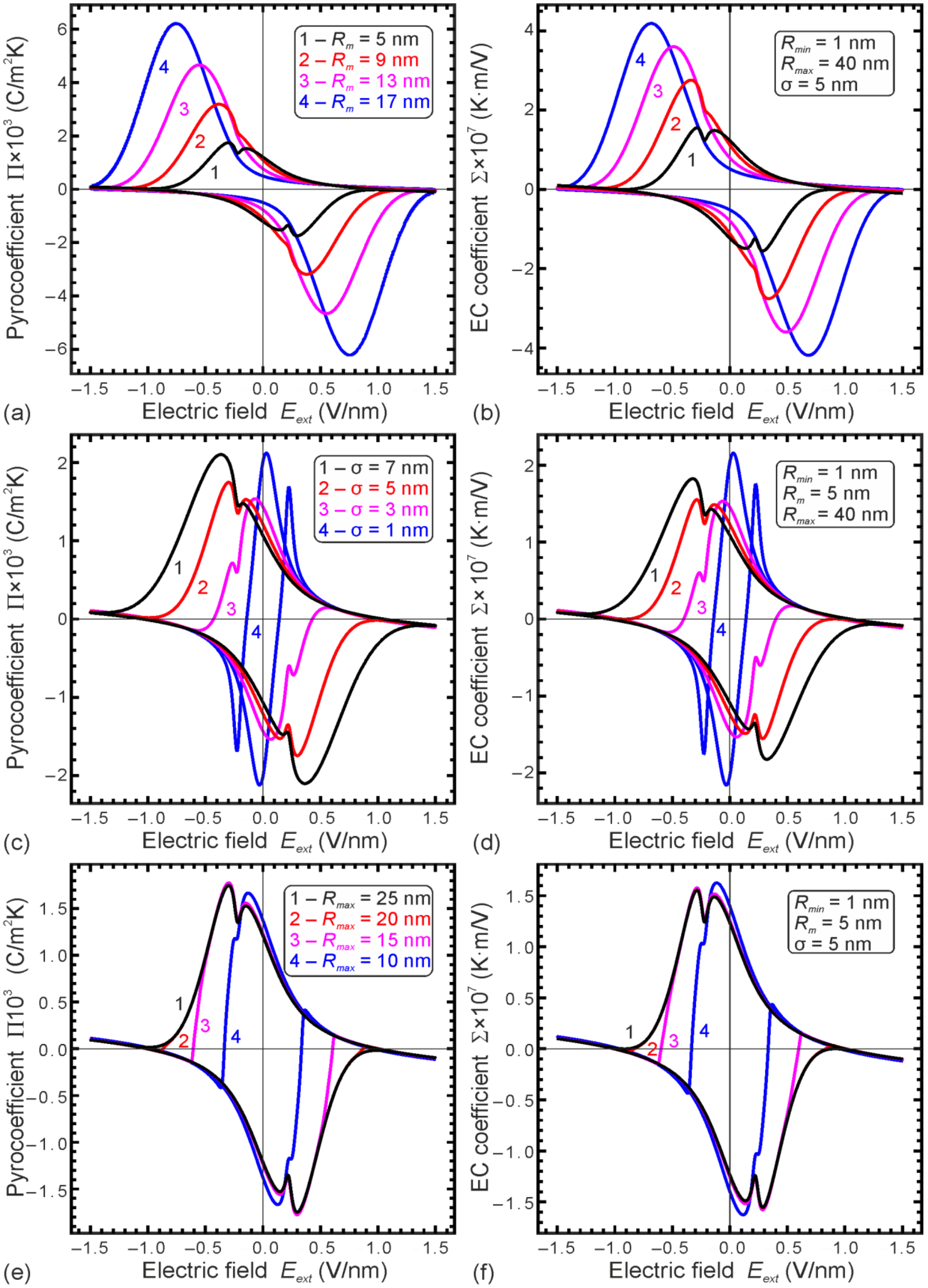

**FIGURE 5.** Dependences of the PE **(a, c, e)** and EC **(b, d, f)** coefficients on external electric field calculated for an ensemble of noninteracting $BaTiO_3$ nanoparticles (curves 1-4), distributed in accordance with **Figs.2a, 3a and 4a**, respectively (see labels at the plots). Other parameters are the same as in **Fig. 2.**

Field dependences $\Pi(E_{ext})$ and $\Sigma(E_{ext})$ shown in **Figs. 5c, d** correspond to different $\sigma$ in the distribution function $f(R)$, which varies in the interval $1nm \leq \sigma \leq 7\ nm$, fixed $R_{min} = 1\ nm$, $R_m = 5\ nm$ and $R_{max} = 40\ nm$ (**Fig. 3a**). Since $R_m$ is smaller than the critical radius, some of the $\Pi$ and $\Sigma$ loops are characterized by the presence of two positive and two negative maxima corresponding to positive and negative electric field. Other loops have only two maxima, one for the positive, and another one for the negative external field. The shape of the loop for $\sigma = 1\ nm$ is significantly different from the shape of the loops for $\sigma = (3-7)\ nm$. A decrease in $\sigma$ (as well as a decrease of $R_m$) corresponds to a decrease in the height of the one maximum, its "splitting" into 2 maxima, and then to an increase in the height of the other maximum (compare curves 1 – 4 in different plots). Note the shift of the maxima towards higher fields with $R_m$ (or $\sigma$) increase. The origin of the maxima splitting is the increasing contribution from the nanoparticle with radius less than critical.

Hysteresis loops $\Pi(E_{ext})$ and $\Sigma(E_{ext})$ shown in **Figs. 5e, f** correspond to different $R_{max}$ in the distribution function $f(R)$, varying in the interval $10\ nm \leq R_{max} \leq 25\ nm$, and fixed $R_{min} = 1\ nm$, $R_m = 5\ nm$ and $\sigma = 5\ nm$, (**Fig. 4a**). An increase in $R_{max}$ leads to the shift and splitting of the $\Pi$ and $\Sigma$ maxima, which is associated with a decrease in the fraction of small nanoparticles with $R < R_{cr}$ for parameters $R_{min} = 1$ nm, $R_m = 5$ nm and $\sigma = 5$ nm. For instance, on curves 4, the splitting of the maxima has already begun, and the two maxima (for each E-sign) become clear for curves 1.

### C. Nanocomposite Figures of Merit

In Ref.[22], the following functions were considered for nanoparticles (NP) in the form:

$$F_I = \frac{\Pi}{c_{NP}},\ F_f = \frac{\Pi}{\varepsilon_0 \varepsilon_{NP}},\ K_{PE} = \frac{\Pi^2}{\varepsilon_0 \varepsilon_{NP} c_{NP}},\ F_{EQ} = \frac{\Pi^2}{\varepsilon_0 \varepsilon_{NP}},\ F_{EU} = \frac{\Pi^2}{\varepsilon_0 \varepsilon_{NP} c_{NP}^2}. \qquad (12)$$

The absolute values of the functions $F_I$, $F_f$ correspond to the pyroelectric figures of merit (FoM) in the radiation detector mode [1, 6, 31, 32, 33], the absolute values of functions $F_{EQ}$, $F_{EU}$ are the pyroelectric FoM in the energy conversion mode [6, 34], and the function $K_{PE}$ is the pyroelectric coupling constant [1, 33, 34]. For the theoretical study, not only the amplitude, but also the sign of the functions (12) are important.

In functions (12), the PE coefficient $\Pi$ and permittivity of nanoparticles $\varepsilon_{NP}$, as well as their bulk heat capacity $c_{NP} = \rho C_P^{NP}$, are size-dependent [22]. This is due to the dependence of the critical transition temperature $T_{cr}$ between the ferroelectric and paraelectric phases on the size $R$ of the nanoparticles [see Eqs. (1), (2) and (4)]. The numerical estimates of FoM for barium titanate are given in the endnote [35].

In order to show the effect of nanoparticle size on FoM, we build dependences (12) on the average particle radius, the expression for which is given by Eq.(2c). **Figures 6–8** show the size dependences of the values (12) calculated for different averaged radii of nanoparticles $\bar{R}$, dispersion $\sigma$, and external electric field $E_{ext}$.

The $\bar{R}$ - dependences of $\delta C_P$, $F_I$, $F_f$, $K_{\mathrm{PE}}$, $F_{EQ}$ and $F_{EU}$, obtained at a weak $E_{ext} = 0.01\,\mathrm{V/nm} \ll E_c$, are shown in **Figs. 6a-d**. This case corresponds to the same values $R_{min} = 1\,\mathrm{nm}$ and $R_{max} = 40\,\mathrm{nm}$, and different values of $\sigma = 7, 5, 4, 3$ nm — curves 1–4, with a change in the shape of the distribution function $f(R)$ specified by the change in the $R_m$ value. A decrease in $\sigma$ leads to a narrowing and an increase in the maxima of the $\delta C_P$, $F_I$, $F_f$, $K_{\mathrm{PE}}$, $F_{EQ}$ and $F_{EU}$, and in accordance with the $f(R)$ change (**Fig. 3a**). The shift of these maxima to the smaller $\bar{R}$, the largest for $F_f$, taking into account the dependence of $\bar{R}$ on $f(R)$, can be associated with the deformation of the distribution curve $f(R)$ with a $\sigma$ change (**Fig. 3a**). It is worth noting, that with a decrease in $\sigma$, the position of the maxima $\bar{R}$ approaches $\bar{R} = R_{cr} = 8$ nm [22].

The dimensional dependences of $\delta C_P$, $F_I$, $F_f$, $K_{PE}$, $F_{EQ}$ and $F_{EU}$, on the average radius $\bar{R}$, obtained for various $E_{ext}$ from $0.01\,\mathrm{V/nm} \ll E_c$ to $1\,\mathrm{V/nm} > E_c$ are shown in **Fig. 7**. An increase in $E_{ext}$ leads to a shift in the maxima of the $\delta C_P(\bar{R})$, $K_{PE}(\bar{R})$, $F_{EQ}(\bar{R})$ and $F_{EU}(\bar{R})$ toward a smaller $\bar{R}$. This displacement is associated with the deformation of the distribution curve $f(R)$ with decreasing $R_m$ and a given $\sigma$ (**Fig. 2a**) and / or decreasing $\sigma$ and a given $R_m$ (**Fig. 3a).** Thus, the action of a weak E-field ( $E_{ext} \ll E_c$ ) is to some extent equivalent to a change in $R_m$ and $\sigma$. The strong E-field $(E_{ext} \sim E_c)$ destroys the maxima.

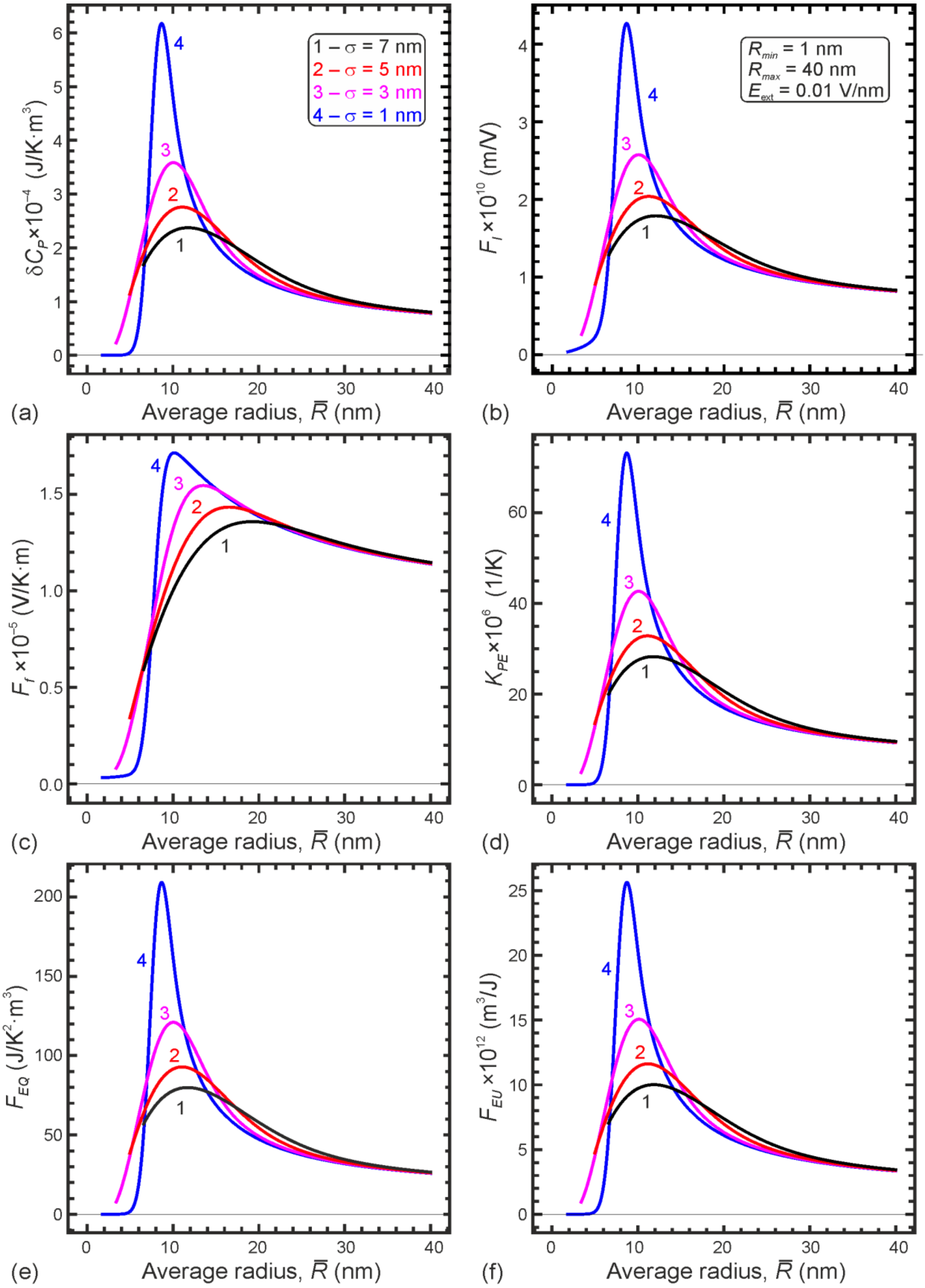

**FIGURE 6.** Dependences of specific heat variation $\delta C_P \equiv C_P - C_P^0$ **(a),** and PE performances $F_I$ **(b),** $F_f$ **(c),** $K_{PE}$ **(d),** $F_{EQ}$ **(e)** and $F_{EU}$ **(f)** on average radius $\bar{R}$ calculated for different dispersions $\sigma =$ 7, 5, 3, 1 nm of $BaTiO_3$ nanoparticles (curves 1-4) in the ensemble, $T = 293\,\mathrm{K}$, $E_{ext} = 0.01\,\mathrm{V/nm}$. Other parameters are the same as in **Fig. 3.**

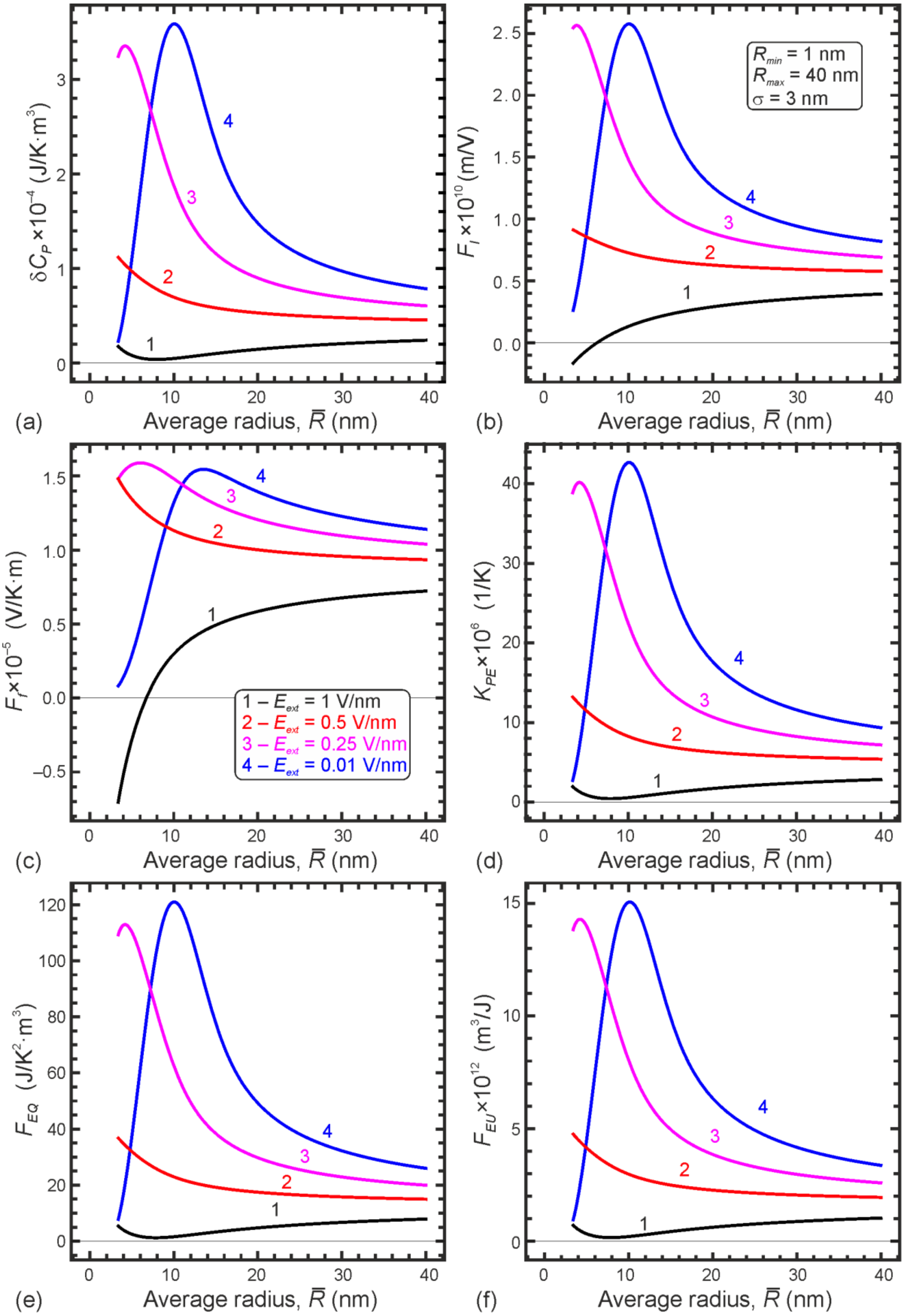

**FIGURE 7.** Dependences of specific heat variation $\delta C_P \equiv C_P - C_P^0$ **(a),** and PE performances $F_I$ **(b),** $F_f$ **(c),** $K_{PE}$ **(d),** $F_{EQ}$ **(e),** and $F_{EU}$ **(f)** on the average radius $\bar{R}$, calculated for different external electrical fields $E_{ext}$ = 1, 0.5, 0.25, 0.01 V/nm of $BaTiO_3$ nanoparticles (curves 1-4) in the ensemble, $T = 293\,\mathrm{K}$, $\sigma = 3\,\mathrm{nm}$. Other parameters are the same as in **Fig. 3.**

The dependences of $\delta C_P$, $F_I$, $F_f$, $K_{PE}$, $F_{EQ}$ and $F_{EU}$ on the average radius $\bar{R}$, for various values of $\sigma$ obtained at a strong $E_{ext} = 1\,\mathrm{V/nm} > E_c$ are shown in **Figs. 8a-d.** Under the strong $E_{ext}$, the character of the dimensional dependences changes significantly in comparison with the case of a weak external field $E_{ext} = 0.01\,\mathrm{V/nm} << E_c$ (see **Fig. 6**). In contrast to the case of weak $E_{ext}$, the dependences $F_I\left(\bar{R}\right)$ and $F_f\left(\bar{R}\right)$ are smoothed (**Figs. 8b, c**), and the $\delta C_P$, $F_I$, $F_f$, $K_{PE}$, $F_{EQ}$ and $F_{EU}$ maxima are replaced by minima (**Figs. 8d, e, f),** which deepen and shift towards smaller $\bar{R}$ with decreasing $\sigma$. Concurrently, the values $\delta C_P\left(\bar{R}\right)$, $F_I\left(\bar{R}\right)$ and $F_f\left(\bar{R}\right)$, as well as $K_{PE}\left(\bar{R}\right)$, $F_{EQ}\left(\bar{R}\right)$ and $F_{EU}\left(\bar{R}\right)$, change by orders of magnitude (compare **Fig. 6** and **Fig. 8**).

The decrease in $\delta C_P$, $K_{PE}$, $F_{EQ}$ and $F_{EU}$ can be associated with their suppression, and the shift of the PE – FE region, by the phase transition under the action of $E_{ext} > E_c$ (Fig. 6 and Fig. 8 in [22]). It should be noted, that the degree of this suppression is different for particles of different radii due to the size shift of the PE – SE phase transition region (Fig. 8 in [22]). The increase in $F_I\left(\bar{R}\right)$ and $F_f\left(\bar{R}\right)$, apparently, is caused by a different degree of suppression of the quantities in the numerator and in the denominator of $F_I\left(\bar{R}\right)$ and $F_f\left(\bar{R}\right)$ [see Eq.(12)] – greater for $\delta C_P$, $\varepsilon_{NP}$, and less for $\Pi$.

The sign change of $F_I\left(\bar{R}\right)$ (**Fig. 8b**) and $F_f\left(\bar{R}\right)$ (**Fig. 8c**) in the vicinity of $R_{cr}$ can be explained by the difference in the sign of $\Pi = -dP/dT$ in the region of the $E$-field-induced PE-FE phase transition for particles with the most probable radius $R_m$ smaller and larger than the critical radius $R_{cr} = 8\,\mathrm{nm}$ (see **Fig. 3c** in [22]).

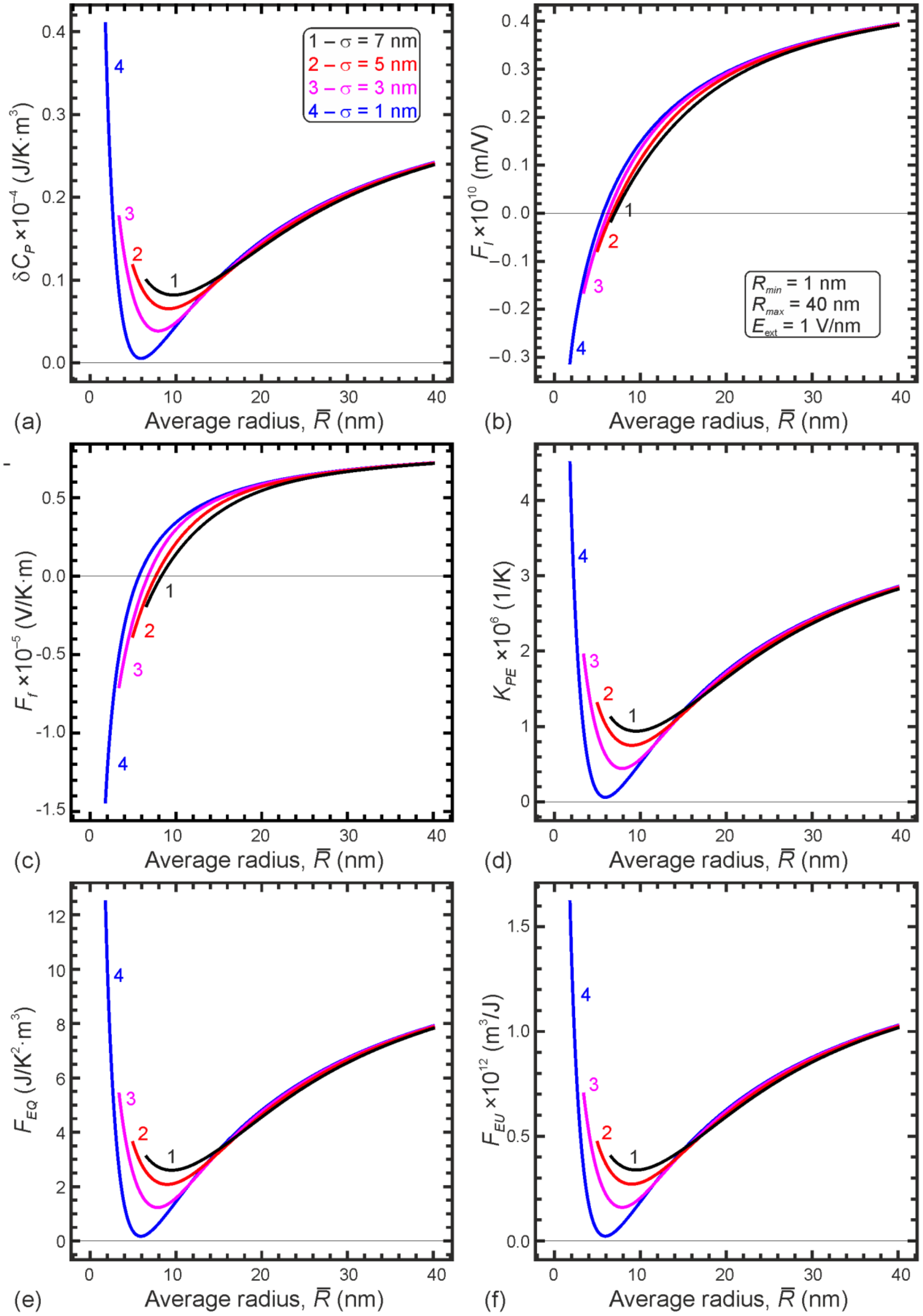

**FIGURE 8.** Dependences of specific heat variation $\delta C_P \equiv C_P - C_P^0$ **(a),** and PE performances $F_I$ **(b),** $F_f$ **(c),** $K_{PE}$ **(d),** $F_{EQ}$ **(e)** and $F_{EU}$ **(f)** on the average radius $\bar{R}$ calculated for different dispersions $\sigma =$ 7, 5, 3, 1 nm of $BaTiO_3$ nanoparticles (curves 1-4) in the ensemble, $T = 293\,K$ , $E_{ext} = 1$ V/nm . Other parameters are the same as in **Fig. 3.**

In summary, for the structure under study, we calculated and analyzed the dependences of the figure of merit on the average particle radius. The characteristics indicate the presence of a phase transition induced by a change in particle size, which is characterized by the presence of a maxima near the critical radius. The value of this radius increases [in the range of (8–12) nm] with an increase in the standard deviation [in the range of (1–7 nm)].

## Conclusion

For noninteracting spherical ferroelectric nanoparticles of various sizes embedded in a dielectric matrix, we calculated the hysteresis loops of polarization EC temperature change, PE and EC coefficient, and dielectric permittivity. We then analyzed the change in the shape of the loops at various values of the Gaussian particle size distribution parameters, namely, the most probable and maximum radii, as well as the mean-square dispersion (as a matter of fact, half-width) of the particle size distribution function.

(a) We have demonstrated that for the same dispersion, the remanent polarization, coercive field, maximums of dielectric permittivity and negative maxima of EC temperature changes strongly depend on the most probable radius, and weakly depend on the maximum radius.

(b) For nanoparticles with the most probable radius $R_m$ less than the critical radius $R_{cr}$ induced by the size of the phase transition at the same minimal and maximal radii, the dielectric permittivity maximums change only slightly, and the remanent polarization, the coercive field and the negative maxima of EC temperature change decrease with a decreasing dispersion of the size distribution function.

(c) In an external electric field much smaller than the coercive field $E_c$ and at constant minimal and maximal radii, the maxima of the size dependences of the pyroelectric figures of merit and the changes in heat capacity in the vicinity of $R \approx R_{cr}$ increase, narrow and shift to smaller average radius values with a decrease in dispersion of the nanoparticle size distribution function.

(d) For nanoparticles with a narrow size distribution, a gradual increase in the external electric field causes a shift in the maxima of the pyroelectric figures of merit and changes in heat capacity to smaller values of the average radius.

(e) A strong electric field ($E_{ext} \sim E_c$) suppress the maxima of the pyroelectric figures of merit. In this case, the dependences of the pyroelectric figures of merit in the radiation detector mode are smoothed out, and the maxima of the pyroelectric figures of merit in the energy conversion mode are replaced by minima that deepen and shift towards smaller values of the average radius as the dispersion of the nanoparticle size distribution decreases.

**Acknowledgements.** A.N.M. expresses her deepest gratitude to the Center for Nanophase Materials Sciences, which is a DOE Office of Science User Facility (CNMS Proposal ID: 257). This project has received funding from the European Union’s Horizon 2020 research and innovation programme under the Marie Skłodowska-Curie grant agreement No 778070.

**Authors' contribution.** A.N.M. and N.V.M. generated the research idea, formulated the problem, and wrote the manuscript draft. H.V.S. wrote the codes and performed the numerical calculations. G.S.S. and V.V.S. worked on the results discussion and the manuscript improvement.

35 **The estimations for $BaTiO_3$ at RT** ($\Pi \approx 2 \cdot 10^{-4}$ C/m$^2$K; $\varepsilon_{NP}\varepsilon_0 \approx 2 \cdot 10^{2} \cdot 10^{-11}$ F/m; $c_{NP} \approx 2 \cdot 10^6$ J/m$^3$K):

$F_I \sim 2 \cdot 10^{-4}$ C/m$^2$K/$2 \cdot 10^6$ J/m$^3$K $\sim 10^{-10}$ m/V = $10^{-10}$ A·m/W = $10^{-10}$ (A/m$^2$)/(W/m$^3$);

$F_f \sim 2 \cdot 10^{-4}$ C/m$^2$K/$2 \cdot 10^{2} \cdot 10^{-11}$ F/m $\sim 10^5$ V/K·m = $10^5$ (V/m)/K;

$K_{PE} \sim 4 \cdot 10^{-8}$ C$^2$/m$^4$K$^2$/$2 \cdot 10^{2} \cdot 10^{-11}$ F/m·$2 \cdot 10^6$ J/m$^3$K $\sim 10^{-5}$ K$^{-1}$;

$F_{EQ} \sim 4 \cdot 10^{-8}$ C$^2$/m$^4$K$^2$/$2 \cdot 10^{2} \cdot 10^{-11}$ F/m $\sim 2 \cdot 10^1$ J/K$^2$·m$^3$ = $2 \cdot 10^1$ (J/m$^3$)/K$^2$;

$F_{EU} \sim 4 \cdot 10^{-8}$ C$^2$/m$^4$K$^2$/$2 \cdot 10^{3} \cdot 10^{-11}$ F/m·$4 \cdot 10^{12}$ J$^2$/m$^6$K$^2$ $\sim 0.5 \cdot 10^{-12}$ m$^3$/J.